\documentclass[ reprint, amsmath,amssymb, aps,pra,
floatfix]{revtex4-2}

\usepackage{graphicx}
\usepackage{dcolumn}
\usepackage{bm}
\usepackage[dvipsnames]{xcolor}
\newcommand{\p}{\partial}

\newcommand{\nn}{\nonumber}
\newcommand{\ta}{\theta}

\newcommand{\be}{\begin{equation}}                                       
\newcommand{\ee}{\end{equation}}
\newcommand{\ba}{\begin{eqnarray}}
\newcommand{\ea}{\end{eqnarray}}
\newcommand{\bl}{\begin{align}}
\newcommand{\el}{\end{align}}

\begin{document}

\preprint{APS/123-QED}

\title{Walk-off induced dissipative breathers and dissipative breather gas in microresonators}

\author{A. Villois$^{1,4}$}
\author{D.N. Puzyrev$^{2,3}$}
 \author{D.V. Skryabin$^{2,3}$}
\author{M. Onorato$^{1}$}
 \affiliation{$^{1}$Dipartimento di Fisica, Università degli Studi di Torino and INFN, 10125 Torino, Italy}
 \affiliation{$^{2}$Department of Physics, University of Bath, Bath, BA2 7AY, UK}
 \affiliation{$^{3}$Centre for Photonics and Photonic Materials, University of Bath, Bath, BA2 7AY, UK}
 \affiliation{$^{4}$School of Mathematics, University of East Anglia,
Norwich Research Park, Norwich, NR4 7TJ, United Kingdom}
\date{\today}

\begin{abstract}
Dissipative solitons in optical microcavities have attracted significant attention in recent years due to their direct association with the generation of optical frequency combs. Here, we address the problem of dissipative soliton breathers in a microresonator with second-order nonlinearity, operating at the exact phase-matching for efficient second-harmonic generation. We elucidate the vital role played by the group velocity difference between the first and second harmonic pulses for the breather existence. We report the dissipative breather gas phenomenon, when multiple breathers propagate randomly in the resonator and collide nearly elastically. Finally, when the breather gas reaches an out-of-equilibrium statistical stationarity, we show how the velocity locking between first and second harmonic is still preserved, naming such phenomena \emph{turbulence locking}.
\end{abstract}

\maketitle

\section{Introduction}
Ultra-high Q-factor optical cavities, such as whispering gallery modes (WGMs) microresonators, offer a compact and power-efficient platform for generating various types of spectrally broadband waveforms, including dissipative Kerr solitons~\cite{Herr2014TemporalMicroresonators,Fulop2018High-orderMicroresonators}, Turing patterns~\cite{HuangGloballyOn-Chip}, soliton molecules~\cite{Weng2020HeteronuclearMicroresonators}, and soliton crystals~\cite{Karpov2019DynamicsMicroresonators}. Spectra of  these waveforms are associated with optical frequency combs. 
Recent developments have shown that a similar range of effects happens  in microresonators with the second-order,  $\chi^{(2)}$, nonlinearity~\cite{Obrzud2018FrequencyMicroresonators, Szabados2020FrequencyMicroresonators,Villois2019FrequencyOscillator.,Amiune2023Mid-infraredMicroresonators,Puzyrev2021Bright-solitonMicroresonators}. Coherent and equidistant microresonator combs can be used for various applications such as, e.g., precision spectroscopy, optical clocks, and search for exoplanets~\cite{Beloy2021FrequencyNetwork,Suh2018SearchingAstrocomb,Kippenberg2018DissipativeMicroresonators}.\\
One of the prerequisites for the efficient operation of a soliton-based microresonator device is the stability of dissipative solitons. Hence, the accurate knowledge of the parameters responsible for instabilities is a fundamental problem relevant to technological applications. On the other hand, understanding soliton instabilities enable the experimental realisation of breather combs~\cite{Guo2017IntermodeMicroresonators,Afridi2022BreatherMicroresonators,Yu2017BreatherMicroresonators} and also opens up the possibility of using microresonators as a platform to investigate the interplay between dissipative solitons, multi-mode chaos and turbulence ~\cite{Anderson2022DissipativeCavities,Chen2020Chaos-assistedMicrocombs,Coulibaly2019Turbulence-InducedResonators}. The study of dissipative breathers in $\chi^{(2)}$ microresonators has so far been restricted to the regime of optical parametric down conversion~\cite{Skryabin1999InstabilitiesOscillators,Parra-Rivas2022DissipativeStability}, where dissipative solitons become Hopf unstable leading to the breather formation. This work will investigate whether this scenario takes place as well in a $\chi^{(2)}$ microresonator set for second harmonic generation (SHG). \\
Assuming that the second-harmonic is far from the phase-matching conditions, the nonlinear interactions resemble the four-wave-mixing (cascading regime), and one can expect the existence of dissipative breathers similar to those observed in Kerr microresonators. However, it remains an open question whether dissipative breathers can also exist at the exact phase-matching (SHG breathers). This question is not limited to microresonators, but it also extends to bulk crystals. Recent studies have demonstrated breather solutions only in the cascading regime,
and brought up close analogies with the Ahkmediev and Kuznetsov-Ma breathers
\cite{Baronio2017AkhmedievMedium,Kuznetsov1977SolitonsPlasma,Ma1979TheEquation,Akhmediev1986ModulationEquation}.
In this work, we show that dissipative solitons at the phase matching point (SHG solitons) are linearly stable, provided that the group velocity walk-off is zero, and, therefore, SHG breathers do not exist. Inducing a non-zero walk-off, which is practically unavoidable in experiments, opens up a window of instability not only for the continuous-wave (cw) solution~\cite{Leo2016Walk-Off-InducedGeneration}, but also for the solitons and, consequently, triggers SHG breathers.
In addition, our findings give evidence that the walk-off-induced instability can also lead  
to the generation of multiple randomly moving and quasi-elastically interacting dissipative breathers. We will refer to such a state as SHG dissipative breather
gas and we will compare it with its SHG dissipative soliton
gas counterpart. Furthermore, we find that when the light inside the resonator reaches a statistically stationary turbulent state, the locking between the first and second-harmonic components of the chaotically moving, disappearing, and emerging pulses remains intact. We refer to this phenomenon as the \emph{turbulence-locking} regime.
\section{Model}
In this work, we will consider a WGM LiNbO$_3$ ring microresonator as the one 
recently used to experimentally generate second harmonic frequency 
comb~\cite{Szabados2020FrequencyMicroresonators}. 
By pumping ordinary polarised light with cw-laser at 1065 nm, it is possible to achieve natural phase matching and generate extraordinary polarised second harmonic light. 
Dispersion relations around the first harmonic cavity mode $m_p$ and second harmonics cavity mode $m_s=2m_p$ are defined as 
\begin{eqnarray}
     &&\omega^{e}_{\mu}=\omega_{m_p}+\mu D_{1p}+\frac{1}{2}\mu^2D_{2p}\\
     &&\omega^o_{\mu}=\omega_{m_s}+\mu D_{1s}+\frac{1}{2}\mu^2 D_{2s}.
\end{eqnarray}
Here $\omega_{m_p}$ and $\omega_{m_s}$ denote the cold cavity resonant frequencies, respectively, while $\mu=0,\pm1,\pm2,\dots$ represents the mode number offset with respect to $m_{p,s}$. The group velocity dispersion coefficients $D_{2p}/2\pi = -100$ kHz and $D_{2s}/2\pi = -200$ kHz are both normal, while the repetition rate is $D_{1p}/2\pi = 21$ GHz. The walk-off parameter $\mathcal{U}=(D_{1p}-D_{1s})$ is of the order of 1 GHz at the phase matching point, $2\omega_{m_p}=\omega_{m_s}$. Throughout this work we will treat the walk-off as a free parameter.
The equations governing the evolution of the first and second harmonic envelops $\psi_{p,s}$ are given by
\ba
&&\nn i\p_t\psi_p=\left(\delta_p-iD_{1p}\p_\ta-\tfrac{1}{2}D_{2p}\p^2_\ta\right)\psi_p-i\frac{\kappa_p}{2}\psi_p+h\\  &&  -\gamma_{p}\psi_s\psi_p^{*},\label{e1}\\
&& \nn i\p_t\psi_s=\left(\delta_s-iD_{1s}\p_\ta-\tfrac{1}{2}D_{2s}\p^2_\ta\right)\psi_s-i\frac{\kappa_s}{2}\psi_s\\  && -\gamma_{s}\psi_p^2  ,\label{e2}\ea

formal derivation can be found in~\cite{Skryabin2020Coupled-modeNonlinearity}.
Here, $\delta_p=\omega_{m_p}-\Omega$ represents the detuning of the cw-pump laser frequency $\Omega$ from the cold cavity resonance frequency $\omega_p$. Similarly, $\delta_s=\omega_{m_s}-2\Omega$ denotes the detuning of the second harmonic. From the detuning definitions it is possible to rewrite the second harmonic detuning as $\delta_s=2\delta_p-\varepsilon$ where $\varepsilon=2\omega_{m_p}-\omega_{m_s}$ is the frequency mismatching parameter. The linewidths are given by $\kappa_p/2\pi=1$ MHz and $\kappa_s/2\pi=4$ MHz, setting the resonator finesse $F=D_{1p}/\kappa_p\sim 10^4$. The nonlinear coefficients are defined as $\gamma_{p,s}/2\pi=300$ MHz $\text{W}^{-1/2}$. The power of the cw-laser $\mathcal{W}$ is related to the pump parameter $h=i\kappa_p/2 \sqrt{\eta/\pi F \mathcal{W}}$, where $\eta$ is the coupling coefficient, $\eta=0.5$.\\

\section{Quadratic dissipative solitons and walk-off-induced instability}
One of the main effects that the walk-off has on dissipative quadratic solitons is the fact that localised structures cannot travel with linear group velocity, but their velocity is selected by the dissipative effects. The value of the velocity at which the first and the second harmonic solitons are locked together is such that the linear momentum $M$ is conserved in time, i.e.,
\begin{equation}\label{Eq:consM}
\frac{d}{dt} M= - (\kappa_p M_p +\kappa_s M_s) =0,
\end{equation}
where 
\begin{eqnarray}
&& M=M_p+M_s\\ \label{eq:M_p}
&& M_p=\gamma_s/i\int_{-\pi}^{\pi} \left( \psi_p^*\partial_{\theta}\psi_p-c.c\right)d\theta\\ 
&& M_s=\gamma_p/(2i)\int_{-\pi}^{\pi} \left( \psi_s^*\partial_{\theta}\psi_s-c.c\right)d\theta.\label{eq:M_s}
\end{eqnarray}
The functional form of ~\eqref{eq:M_p} and~\eqref{eq:M_s} is such that $M_p+M_s$ is conserved in time in the Hamiltonian limit, $\kappa_{p,s} = 0$, see~\cite{Skryabin2001WalkingSolitons}. 
Note that the momentum is not, in general, a conserved quantity, given the presence of dissipation, however,  the system can support travelling wave solutions moving with the common velocity $\mathcal{V}$ if the rhs of Eq.~\eqref{Eq:consM} vanishes.\\
A natural question that arises is how to estimate the value of such velocity. In order to answer this, 
it is instructive to consider the case when $\kappa_{p,s} = 0$. In
such Hamiltonian limit, soliton solutions are not limited to a single value of velocity locking, but $\mathcal{V}$ can span a continuous range of values~\cite{Skryabin2001WalkingSolitons}.
Since an analytical solution for second harmonic $\chi^2$ cavity solitons does not exist, to our knowledge, it is useful to start identifying the range of possible velocities at which a solitary travelling wave can propagate in microresonator. To do so, one can relies on the so-called band gap analysis already introduced in ~\cite{Villois2019SolitonMicroresonators}.\\
The idea behind the band gap analysis is the following: a localised/solitary waves moving with velocity $\mathcal{V}$ in an optical cavity can exist on the top of a homogeneous cw-solution $\psi^0_{p,s}$ only if the condition
\begin{equation}\label{Eq:res}
  \mathcal{V}\neq \frac{\omega_{\mu}}{\mu}
\end{equation}
is true for each $\mu$ in the cavity, where $\omega_{\mu}$ represents the dispersion relation for weakly nonlinear waves propagating on the top of the cw-solution. 
Such dispersion relation can be found performing standard Bogoliubov-de Gennes analysis. In order to emphasise the role of the walk-off parameter, we will work in a frame of reference moving  with angular velocity $D_{1p}$.
One can now substitute the ansatz
\begin{equation}
    \psi_{p,s}=\psi^0_{p,s}+u_{p,s} e^{i\mu\theta+\lambda t }+v^{*}_{p,s} e^{-i\mu\theta+\lambda^* t }
\end{equation} with $\lambda\in \mathbb{C}$,
in Eqs.~\eqref{e1} and Eqs.~\eqref{e2} and solve the eigenvalue problem 
\begin{equation}
    \mathcal{M}\mathbf{A}=\lambda\mathbf{A}
\end{equation}
arising after linearising with respect to the small amplitude perturbations $|u_{p,s}|$ and $|v_{p,s}|$, where $\mathbf{A}=(u_p,v_p,u_s,v_s)^T$ and 
\begin{equation}\label{matrix1}
\footnotesize
    \mathcal{M}=
    \begin{bmatrix}
\mathcal{L}_p-i\kappa_p  && -\gamma_p \psi^0_s &&\gamma_p(\psi_p^0)^* && 0 \\
\gamma_p(\psi_s^0)^*&&-\mathcal{L}^*_p-i\kappa_p    && 0 &&  \gamma_p\psi^0 \\
-2g_s\psi^0              && 0              && \mathcal{L}_s-i\kappa_s          && 0      \\
 0             &&2g_s (\psi_p^0)^*        && 0     &&   -\mathcal{L}^*_s-i\kappa_s        \\
\end{bmatrix},
\end{equation}
where $\kappa_{p,s} = 0$ and with the following definitions:
$\mathcal{L}_p\equiv(\delta_p+1/2 D_{2p} \mu^2 )$ and $\mathcal{L}_s\equiv(\delta_s+1/2 d_{2s}\mu^2+\mathcal{U}k)$.
\begin{figure}
    \centering
    \includegraphics[width=0.49\textwidth]{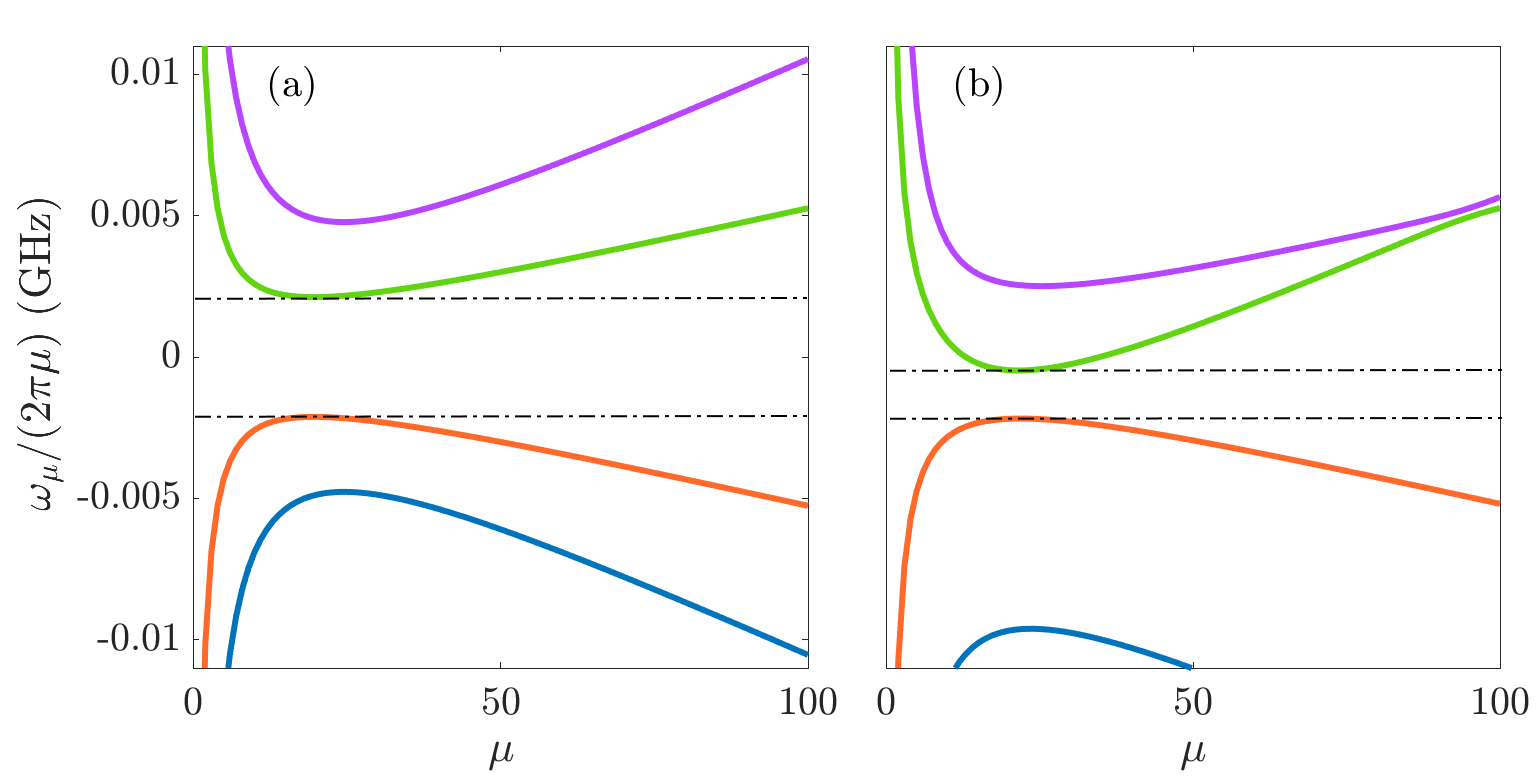}
    \caption{Numerically evaluated phase speed for small amplitude waves on the top lower branch cw-solution obtained for the following parameters: $\delta_p/2\pi=-26.64$ MHz, $\delta_s=2\delta_p$ and $\mathcal{W}=260$ mW. In panel (a) walk-off parameter $\mathcal{U}$ is zero, while in (b) $\mathcal{U}/2\pi=-5$ MHz}
    \label{Fig:1}
\end{figure}
Figure~\ref{Fig:1} shows the four phase speed branches $\omega_{\mu}/\mu$ for low amplitude waves numerically computed in the frame of reference moving with angular velocity $D_{1p}$. Note that $\omega_{\mu}$ corresponds to Im$(\lambda)$. Dash lines in the plot highlight a band gap: a range of velocities such that small amplitude waves cannot propagate and, hence, where it possible to prevent energy transferring from solitonic solutions to small amplitude waves.\\
We can now focus on the role of the walk-off parameter in the band gap analysis. One of the main effect of the walk-off is presented in Fig.~\ref{Fig:1}(b) where it is shown how non-zero walk-off in the system modifies the dispersion of linear waves, leading to the narrowing of the band gap. As we increase the walk-off parameter beyond a critical value, $\mathcal{U}_{c}$, the band gap completely closes, thereby impeding the existence of soliton solutions.\\
The presence of walk-off also has a significant impact on the behaviour of SHG solitons, specifically with respect to the velocity at which the first and second harmonic solitons become locked together.
As already mentioned, by reintroducing losses in the system, first and second harmonic solitons will travel at a speed 
such that the rhs of Eq.~\eqref{Eq:consM} vanishes. Such value will be in general selected by the values of dissipation parameters $\kappa_{p,s}$, as discussed in the appendix~\ref{App:A}, and it will lie within the range estimated using band gap analysis. The specific value of velocity locking can be evaluated numerically, with a velocity selective Newton-Raphson
method, see red circles in Fig.~\ref{Fig:M} in the appendix. Alternatively, it can be analytically shown that for small values of walk-off, the locking velocity, $\mathcal{V}$, depends linearly on $\mathcal{U}$:
\begin{equation}\label{Eq:semi_an}
    \mathcal{V}=\alpha\mathcal{U}.
\end{equation}
Derivation of the $\alpha$ coefficient can be found in the appendix, see Eq.~\eqref{Eq:semi_an_appndix}, where a comparison between the velocity locking values obtained numerically and analytically is shown in Fig.~\ref{Fig:3}(b).\\
\begin{figure}
    \centering
    \includegraphics[width=0.48\textwidth]{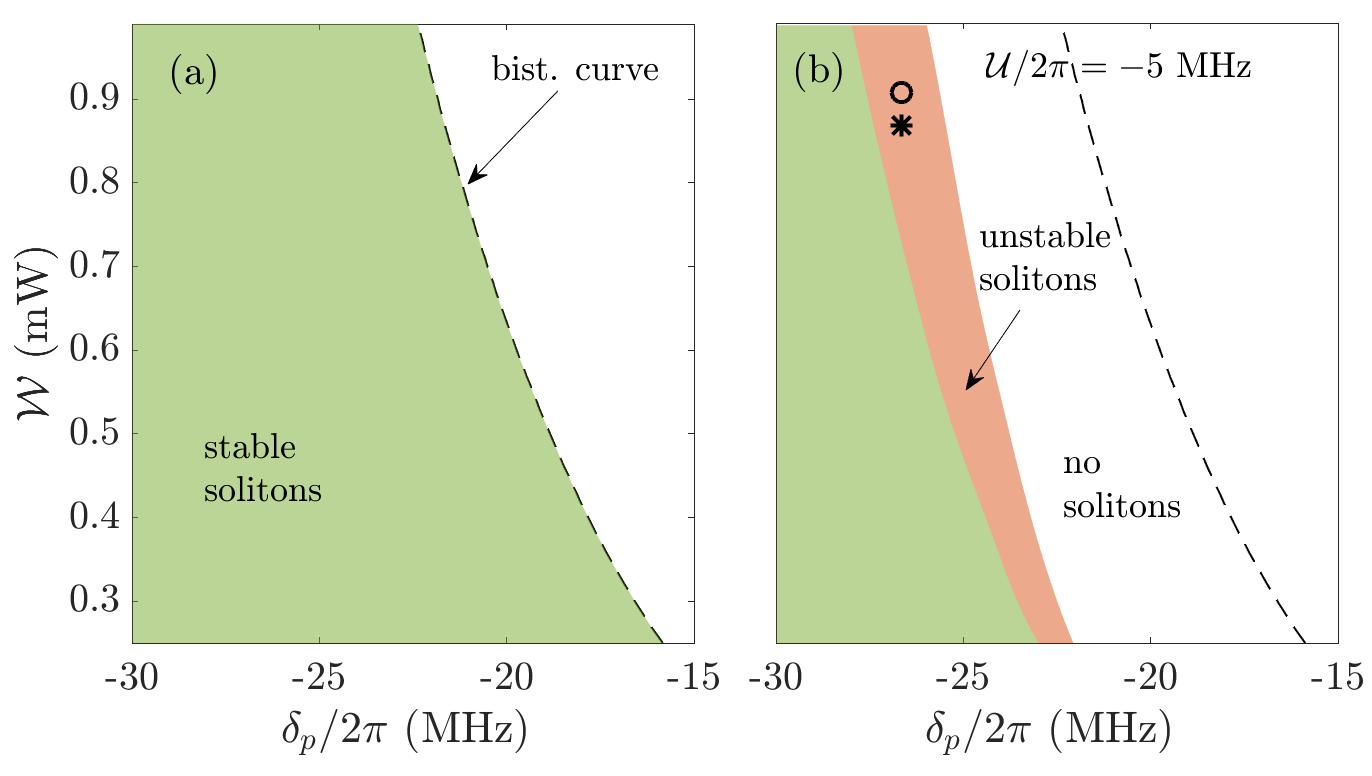}
    \caption{ Existence region for soliton solutions for different values of detuning and cw-laser input power, in the absence of walk-off (a) and with in the presence of walk-off (b). Existence region is studied up to -30 MHz although it can extend beyond such value.
    Stability and instability region are denoted by green (light gray) and orange (dark gray) colour respectively. Bistability line denotes where the homogeneous cw solution becomes multivalued.
    }
    \label{Fig:4}
\end{figure}
\section{Quadratic Dissipative Breathers}
After presenting the influence of the walk-off on the velocity and the existence of SHG soliton solutions, we will now examine its impact on their stability and how it contributes to the transition from solitons to breathers. It is well known that breather
solutions can exist in the integrable NLS equations in the form of time ~\cite{Kuznetsov1977SolitonsPlasma,Ma1979TheEquation} and space~\cite{Akhmediev1986ModulationEquation} periodic solutions. In the case of quadratic nonlinearity, similar solutions have been found only in the cascading regime limit, where four wave mixing is dominant~\cite{Baronio2017AkhmedievMedium}. In our work, such a limit would correspond to the case of $\delta_s\gg \delta_p$. 
For Kerr optical cavities, where driving and dissipations are added to the standard NLS equation (aka Lugiato Lefever equation~\cite{Lugiato1987SpatialSystems}), breathers can exist when solitons undergo a Hopf instability~\cite{Skryabin1999InstabilitiesOscillators}. Numerical and experimental results concerning dissipative breathers can be found in~\cite{Yu2017BreatherMicroresonators,Johansson2019StabilityContinuum}. The main problem concerning finding SHG breathers in the phase-mathcing regime consists in the stability of SHG solitons. 
In order to study linear stability of soliton solutions, one can modify Eq.~\eqref{matrix1} substituting the soliton profile instead of the cw-background solution, considering small amplitude perturbations being generic functions of $\theta$, $u_{p,s}(\theta)$ and $v_{p,s}(\theta)$, and redefining $\mathcal{L}_p\equiv(\delta_p-1/2 D_{2p} \partial^2_{\theta}-i\mathcal{V}\partial_{\theta})$ and $\mathcal{L}_s\equiv(\delta_s-1/2 D_{2s} \partial^2_{\theta}-i(\mathcal{V}-\mathcal{U})\partial_{\theta})$.
Differently from dissipative solitons in the Kerr case, SHG solitons appear to be stable for all values of detuning and/or driving power, see green (light gray) region in Fig.~\ref{Fig:4}(a), as long as the walk-off parameter is neglected. Note that, moving away from the ideal scenario of perfect group velocity matching (zero walk-off)
makes it harder to find SHG solitons, due to the closure of the band gap. Only recently has been proven existence of quadratic dissipative soitons (DSs) with large walk-off, but limited to the cascading regime~\cite{Puzyrev2021Bright-solitonMicroresonators}. However, it is possible to alter the stability of SHG solitons by considering a value of the walk-off parameter large enough to trigger a Hopf instability, see orange (dark gray) region in Fig.~\ref{Fig:4}(b), but less than the critical value, $\mathcal{U}_c$ responsible for the closure of the existence gap.
By selecting detuning or cw-input power within the orange region in Fig.\ref{Fig:4}(b), one can investigate the effects of such instability in the dynamical evolution of the SHG soliton.
\begin{figure*}
    \centering
    \includegraphics[width=0.98\textwidth]{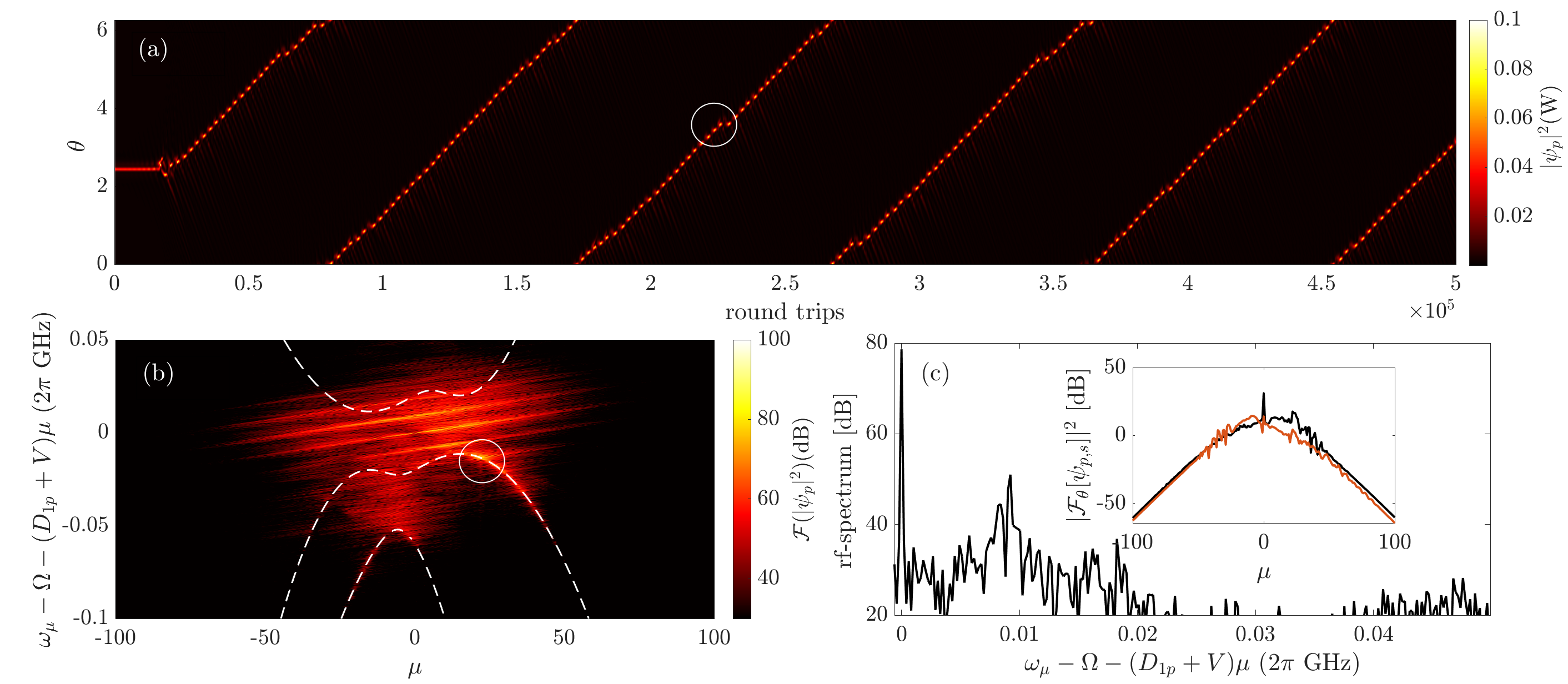}
    \caption{(a) Dynamical evolution of unstable soliton solution. Detuning parameters and cw-laser power are $\delta_p/2\pi=-26.64$ MHz, $\delta_s=2\delta_p$, $\mathcal{W}=870$ mW and $\mathcal{V}=-1.62$ MHz. (b) Space-time Fourier transform of breather dynamics. Space-time Fourier transform within the time interval $[2.5, 2.6]\times 10^6$ round trip times. Weakly nonlinear dispersion relation superimposed using white dash line. (c) rf-spectrum for first harmonic. Inset in (c) shows first and second harmonic spectra, black and orange (gray) line respectively, of the dissipative breathers after $2.6\times 10^6$ round trip times.}
    \label{Fig:5}
\end{figure*}
Figure~\ref{Fig:5}(a) shows how the instability is responsible for the soliton-breather transition. The figure shows the dynamics of the first harmonic component of the SHG soliton in a co-moving frame of reference. This choice was made to emphasise how, after a few thousand of round trip times, the soliton starts radiating dispersive waves and stabilises at a new velocity. The system reaches a meta stable state characterised by random velocity shifts as highlighted by white circles in Fig.~\ref{Fig:5}(a). Meta stability has been tested numerically for up to $4\times10^6$ round trip times.
Given the finite size of the system, the emitted dispersive waves cannot escape but they keep travelling within the cavity. We stress that this is not a numerical artefact, but an intrinsic property of microresonators. This process will trigger a continuous emission and re-absorption of the waves causing the breathing of the soliton~\cite{Skryabin1999InteractionOscillators}. 
The coherence of such state can be observed in the space-time Fourier analysis of the dynamics. Specifically, from Fig.~\ref{Fig:5}(b) it is possible to see how the dissipative breather is formed by the superposition of several coherent structures, all moving with the same velocity. 
The crossing of the breather signal, with the weakly nonlinear dispersion, see white dash line in Fig.~\ref{Fig:5}(b), causes the resonant emission of dispersive waves with characteristic wavenumber $\mu_r\sim22$, see white circle in Fig.~\ref{Fig:5}(b). Note that weakly nonlinear dispersion is the same as the one plotted in Fig.~\ref{Fig:1}(b) but in a frame of reference co-moving with the soliton. Spectra evaluated after $2.6 \times 10^6$ round trip times are shown in the inset of Fig.~\ref{Fig:5}(c) for both first and second harmonic. In the first harmonic spectrum, aside from the main peak associated to the cw-pumping, a smaller secondary peak is also present at $\mu=22$. 
In order to evaluate the oscillation period of the SHG breather, one can evaluate the dominant frequency in the system trough the so-called rf-spectrum, evaluated as $\mathcal{F}_t[\langle \psi |\psi\rangle_{\theta}]$, where $\mathcal{F}_t$ represents the Fourier transform with respect to time and $\langle \psi |\psi\rangle_{\theta}=\int |\psi|^2 d\theta$. Such a plot is shown in Fig.~\ref{Fig:5}(c) where a peak is observed at $0.01$ GHZ. Such value  is remarkably close to half of the gap size, as pointed out in recent literature on dissipative Kerr breathers~\cite{Puzyrev2022FrequencyMicroresonators}.
\section{Dissipative Breather Gas}
The effects of the walk-off induced instability on soliton dynamics can be further studied by examining the transition from a SHG soliton to a turbulent state
 This transition can be achieved by increasing the growth rate, Re$(\lambda)$, of the linearly unstable eigenstates responsible for the soliton instability, for instance, by maintaining the same detuning as shown in Fig.~\ref{Fig:5} while increasing the input cw-power $\mathcal{W}$.
\begin{figure*}
    \centering
    \includegraphics[width=0.98\textwidth]{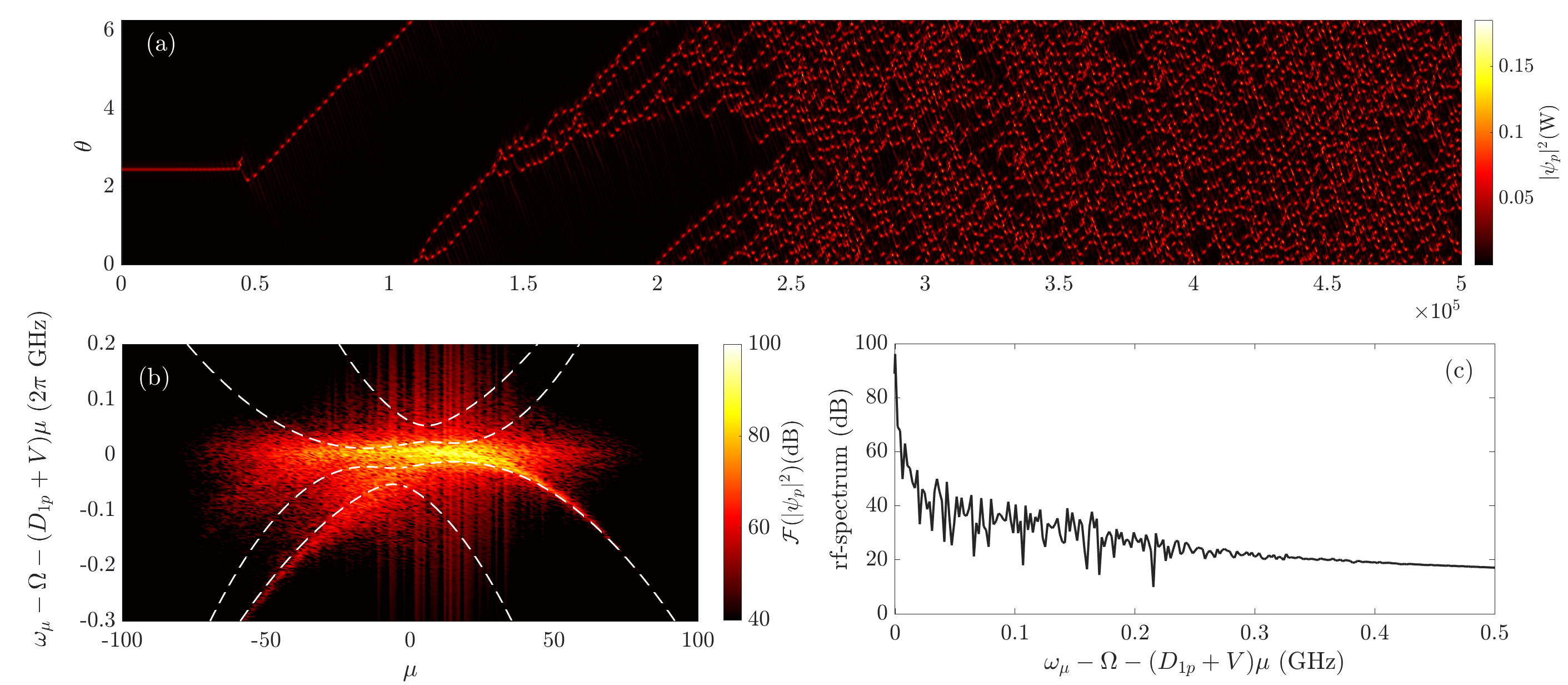}
    \caption{(a) Dynamical evolution of unstable soliton solution. Detuning parameters and cw-laser power are $\delta_p/2\pi=-26.64$ MHz}, $\delta_s=2\delta_p$, $\mathcal{W}=910$ mW and $\mathcal{V}=-1.63$ MHz. (b) Space-time Fourier transform within the time interval $[3, 5] \times 10^4$ round trip times. (c) rf-spectrum evaluated as in Fig.~\ref{Fig:5}(c)
    \label{Fig:6}
\end{figure*}
Figure~\ref{Fig:6}(a) displays the time evolution of the first harmonic component of an unstable SHG soliton inside the resonator. Due to the walk-off induced instability, the soliton first transforms into an SHG breather, then into a state characterized by an increasing number of breathers before finally settling into a turbulent state similar to the Kerr spatiotemporal chaos observed in~\cite{Karpov2019DynamicsMicroresonators}. This state is referred to as SHG \emph{dissipative breather gas}, a term adopted from the integrable turbulence community~\cite{El2020SpectralEquation}. Unlike integrable systems where solitons or breathers can only interact elastically with each other, our system can reach an out-of-equilibrium stationarity characterized by a multitude of pulses continuously emitting dispersive waves. The loss of coherence typical of a turbulent state is evident in both Figs.~\ref{Fig:6}(b) and (c), where there is no longer a clear signal associated with a coherent structure, nor a well-defined peak in the rf-spectrum.\\
A question that arises naturally concerns the difference between SHG dissipative breather and soliton gas.
As already pointed out, differently from dissipative solitons in the Kerr case, SHG solitons are not affected by Hopf-type instability; as long as the walk-off parameter is negligible, this result is independent of the detuning value and the initial power. Therefore, we can exploit this feature to generate a gas of SHG dissipative solitons.
\begin{figure*}
    \centering
    \includegraphics[width=0.98\textwidth]{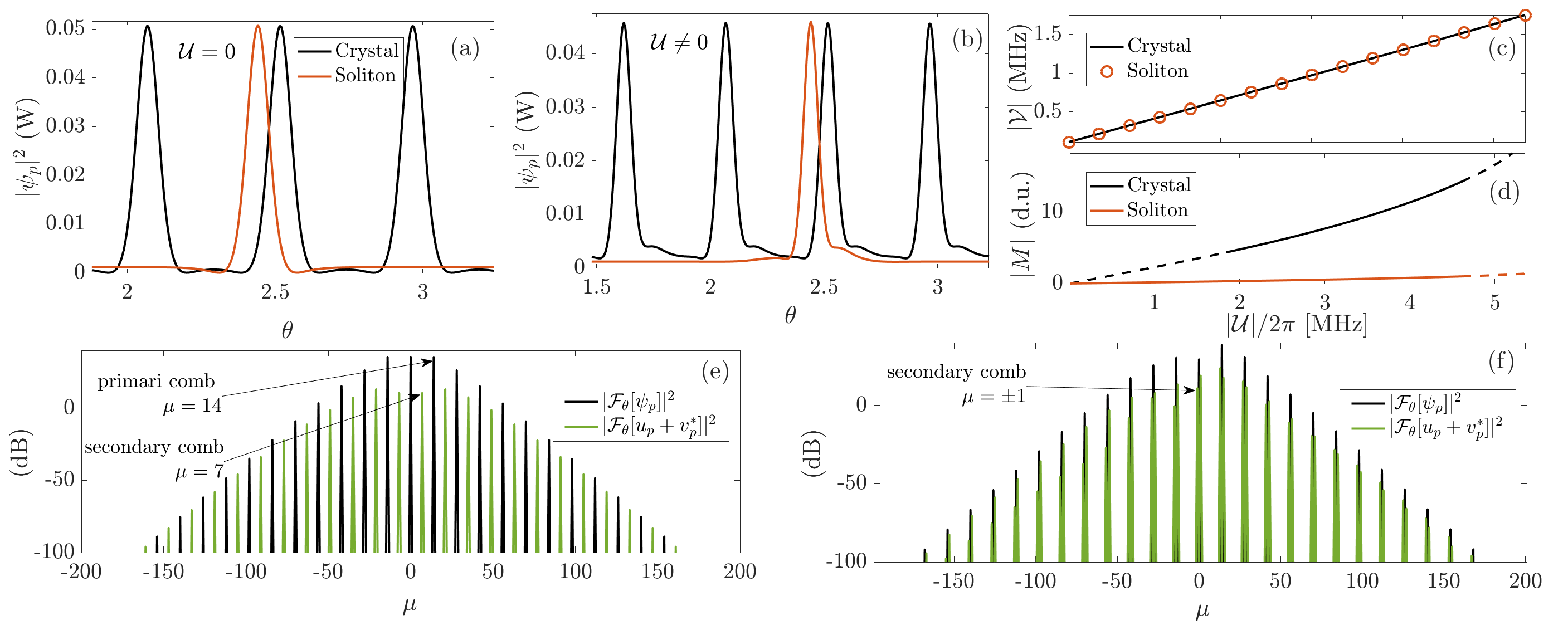}
    \caption{Space profile of stationary DS solution and soliton crystal, orange (gray) and black line respectively, in case of (a) zero walk-off and (b) $\mathcal{U}/2\pi=-5$ MHz. (c) Plot of velocity locking for both single DS and soliton crystal versus walk-off. (d) Total linear momentum in dimensionless units versus walk-off. Dash lines highlight unstable solutions. Fourier spectrum of soliton crystal, black lines, superimposed to Fourier spectrum of unstable crystal eigenstates, green (gray) lines, for (e) zero walk-off and (f) higher walk-off, $|\mathcal{U}|/2\pi=5$ MHz. For each panel detuning and cw-input power are the same as the one used in Fig.~\ref{Fig:6}.}
    \label{Fig:TP}
\end{figure*}
To engineer such a state, we start by considering an initial condition characterised by 15 solitons in both harmonics, which we will refer to it as a soliton crystal. Figures~\ref{Fig:TP}(a) and (b) display a portion of the crystal obtained numerically using a Newton-Raphson method. These figures allow for a visual comparison between the crystal, see black line, and the single soliton solution, see orange (gray) line, for the first harmonic component. Figure~\ref{Fig:TP}(c) demonstrates that the velocity locking of the crystal remains consistent with that of the single SHG soliton solution for different values of the walk-off parameter. Additionally, Fig.~\ref{Fig:TP}(c) confirms the linear trend as predicted in Eq.~\eqref{Eq:semi_an}. A substantial difference between the crystal and the single SHG soliton solution concerns their linear stability. In Figure~\ref{Fig:TP}(d) we plot the dimensionless value of total momentum for the system with respect to different values of walk-off. Despite sharing the same velocity, the crystal exhibits higher momentum due to its higher power intensity. Dash and solid lines in Fig.~\ref{Fig:TP}(d) denotes unstable and stable solutions, respectively.
More specifically, the crystal is unstable for small walk-off values, differently from the single soliton, see black and orange (gray) lines, respectively. By increasing the walk-off value, the crystal stabilizes and then de-stabilizes again, experiencing the same type of walk-off-induced instability experienced by the single SHG soliton which led to the formation of a SHG breather, as discussed in the previous section. 
The different nature between the two instabilities affecting the crystal can be understood by considering the spectrum associated with the most linearly unstable eigenstates of the crystal, see Figs.~\ref{Fig:TP}(e) and (f). In case of a small walk-off, the instability leads to the formation of a secondary comb with different periodicity, while for large walk-off values, the instability leads to the creation of sidebands around the primary comb. 
As a final remark, the instability for small values of walk-off appears in the form of a single positive real eigenvalue $\lambda$, while for large walk-off it appears in the form of a pair of complex conjugate eigenvalues with positive real parts~\cite{Wang2019DissipativeResonators}.\\
We will now study the long term effects of these two different types of instability on the temporal dynamics of the crystal. 
\begin{figure*}
    \centering
    \includegraphics[width=0.98\textwidth]{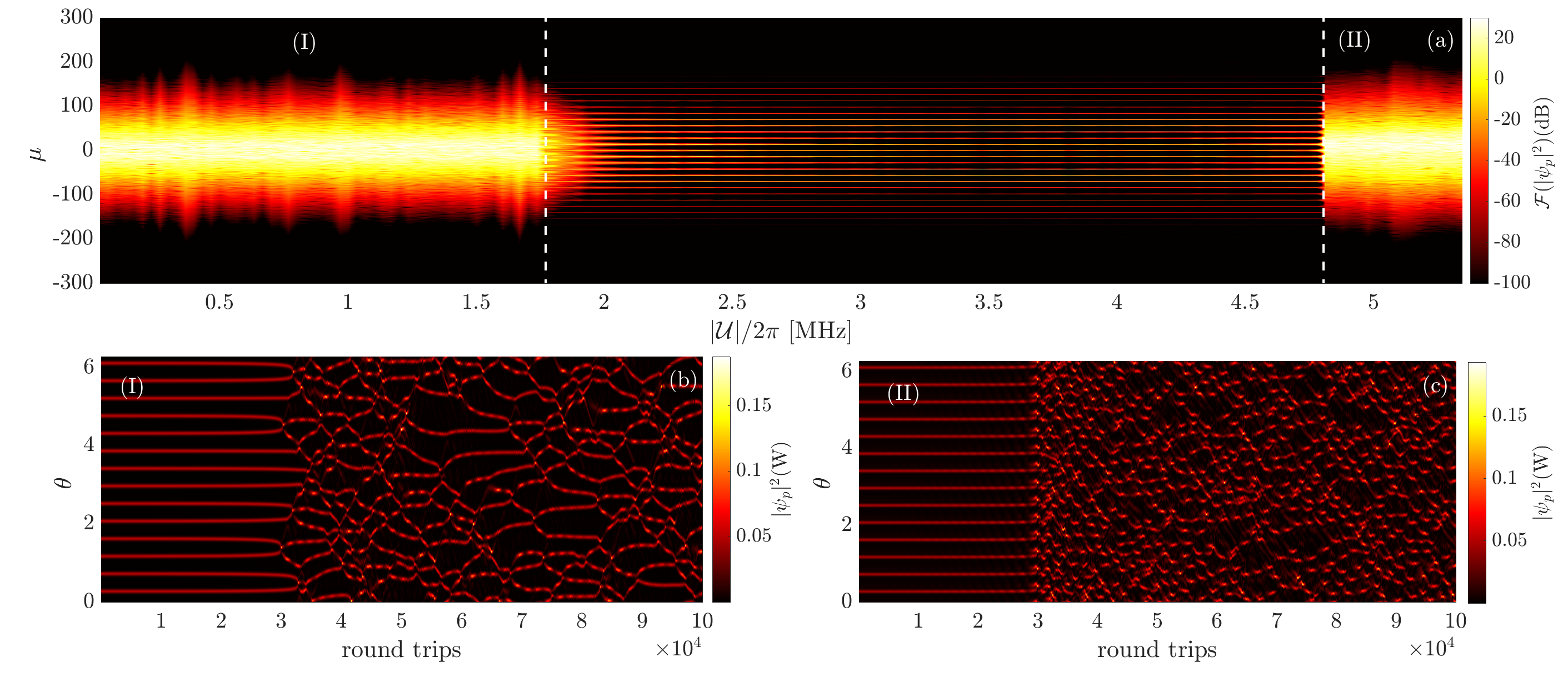}
    \caption{(a) Collection of spectra  from dynamical evolution of stationary crystal states for different values of walk-off parameter. Each spectrum is measured after $5\times10^5$ round trip times. Time evolutions of soliton crystal for (b) $|\mathcal{U}|/2\pi=1$ MHz and (c) $|\mathcal{U}|/2\pi=5$ MHz. For each panel detuning and cw-input power are the same as the one used in Fig.~\ref{Fig:6}. 
    }
    \label{Fig:ws}
\end{figure*}
 Figure~\ref{Fig:ws}(a) shows the first harmonic component of a collection of spectra obtained from the dynamical evolution of stationary crystal states for different values of walk-off parameter. Each spectrum is measured after $5\times10^5$ round trip times. As predicted from the linear stability analysis, see dash lines in Fig.~\ref{Fig:TP}(d), the crystals exhibit instabilities in two separate walk-off regions, denoted by (I) and (II) in Fig.~\ref{Fig:ws}(a). Boundaries of such regions are highlighted by white dash lines. An example of the dynamical evolution of the first harmonic component of the crystal for each unstable region is presented in Figs.~\ref{Fig:ws}(b) and (c), respectively. The instability present in region (I) causes the breaking of the crystal into its individual constituents, which are, however, stable according to the linear stability analysis presented in Fig.~\ref{Fig:4}(a). For this reason, in Fig.~\ref{Fig:ws}(a) it is still possible to identify the trajectories of the individual solitons and their interaction. This type of dynamics is, on contrary, not present in the dynamics associated with region (II), where the crystal seems to transition into a state similar to the one already reported in Fig.~\ref{Fig:6}(a). For this reason, we are led to identify the type of turbulence characterising region (I) as a SHG dissipative soliton gas, while the one characterising region (II) as a SHG dissipative breather gas.
 To compare these two different types of dissipative gases in a more quantitative way, it is possible to study the probability distribution function (PDF) for the field amplitudes. To do this, we repeat the simulations presented in Figs.~\ref{Fig:ws}(b) and (c) considering the same initial conditions but affected by different random noise. 
 \begin{figure}
    \centering
    \includegraphics[width=0.48\textwidth]{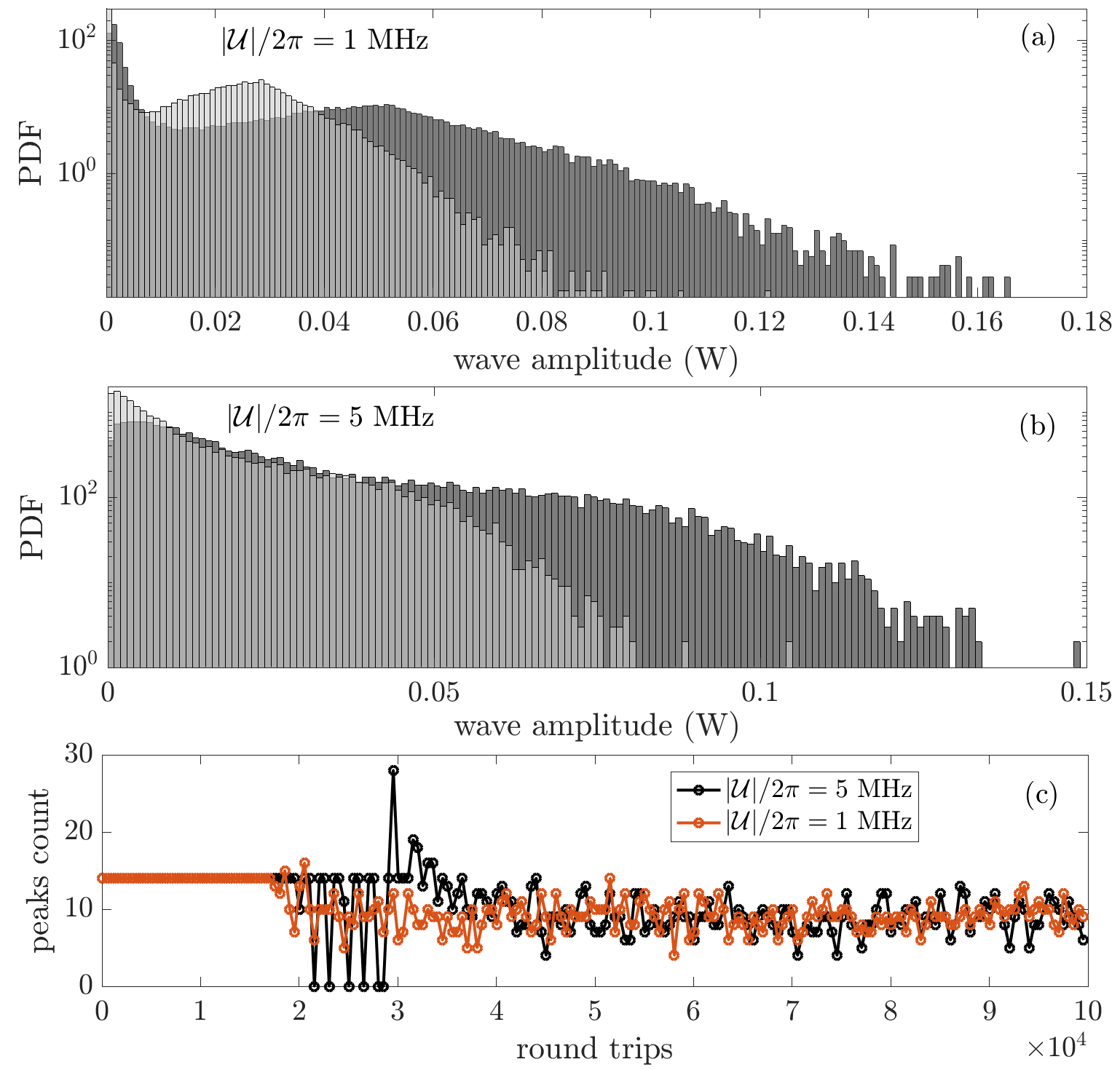}
    \caption{Probability density function or the waves amplitude for many realisation of: (a) dissipative soliton gas, when walk-off is zero; (b) dissipative breather gas, when walk-off is $\mathcal{U}/2\pi=-5$ MHz. Dark gray and light gray histograms are associated to first and second harmonic respectively. (c) Ensemble average of time evolution of first harmonic peak counts higher than 0.04 W. }
    \label{Fig:7}
\end{figure}
To construct a PDF, we follow a process that involves generating a histogram. In this histogram, each bin value, denoted as $v_i$, is determined using the formula $v_i = c_i N/w_i$, where $c_i$ represents the number of elements within the bin, $N$ denotes the total number of data points in the histogram, and $w_i$ represents the bin width, which is set at $1$ mW.
From the PDF associated to region (I), plotted in Fig.~\ref{Fig:7}(a), one can see how the first harmonic has a peak around 0.05 W, while the second harmonic has a peak around 0.03 W, see dark gray and light gray histogram respectively. Such peaks correspond to the amplitude of the single SHG soliton solution, see Fig.~\ref{Fig:TP}(a), justifying the assumption of a SHG dissipative soliton gas.
Note the presence of a peak for low power too, which can be associated to the homogeneous cw-background onto which the solitons lie.
Differently, the PDF for region (II), plotted in Fig.~\ref{Fig:7}(b), appears to be flat for both harmonics. The absence of well defined peaks is due to the multitude of low amplitude dispersive waves constantly emitted by the breathing mechanism and by the fact that each SHG breather has amplitude which oscillates between $|\psi_p|^2\in [0.2,0.8]$ W and $|\psi_s|^2\in [0.1,0.5]$ W for the first and second harmonic, respectively. To make sure that the difference in the PDFs is not simply a consequence of a different number of high amplitude pulses present in the system, one can measure the number of peaks having amplitude higher than a fixed threshold. Figure \ref{Fig:7}(c) displays the temporal evolution of the peak count for the first harmonic dynamics in regions (I) and (II). The peak count is averaged over 25 different realisations characterised by different white noise in the initial condition. 
A threshold value of 0.04 W is selected, slightly lower than the peak power of the single SHG soliton solution. Figure \ref{Fig:7}(c) reveals that during the initial stage of the dynamics, both regions exhibit an equal number of peaks. However, due to the crystals instability, a sudden increase in peak count is observed specifically in the dynamics associated with region (II). This phenomenon can be attributed to the influence of walkoff, resulting in an asymmetric shape within the solitons forming the crystal. This asymmetry is manifested by a minor amplitude bump (as depicted in Fig. 5 (b)) on the right side of each peak within the crystal. Over time, the Hopf instability amplifies this small bump, giving rise to the generation of multiple pulses within the resonator. Consequently, for a brief period, the number of these pulses seems to be twice the count of peaks characterising the crystal.
However, after this transitional phase, both the SHG dissipative breather and soliton gas settle around an average peak count, which becomes comparable.

\section{Turbulence Locking}
\begin{figure}
    \centering
    \includegraphics[width=0.48\textwidth]{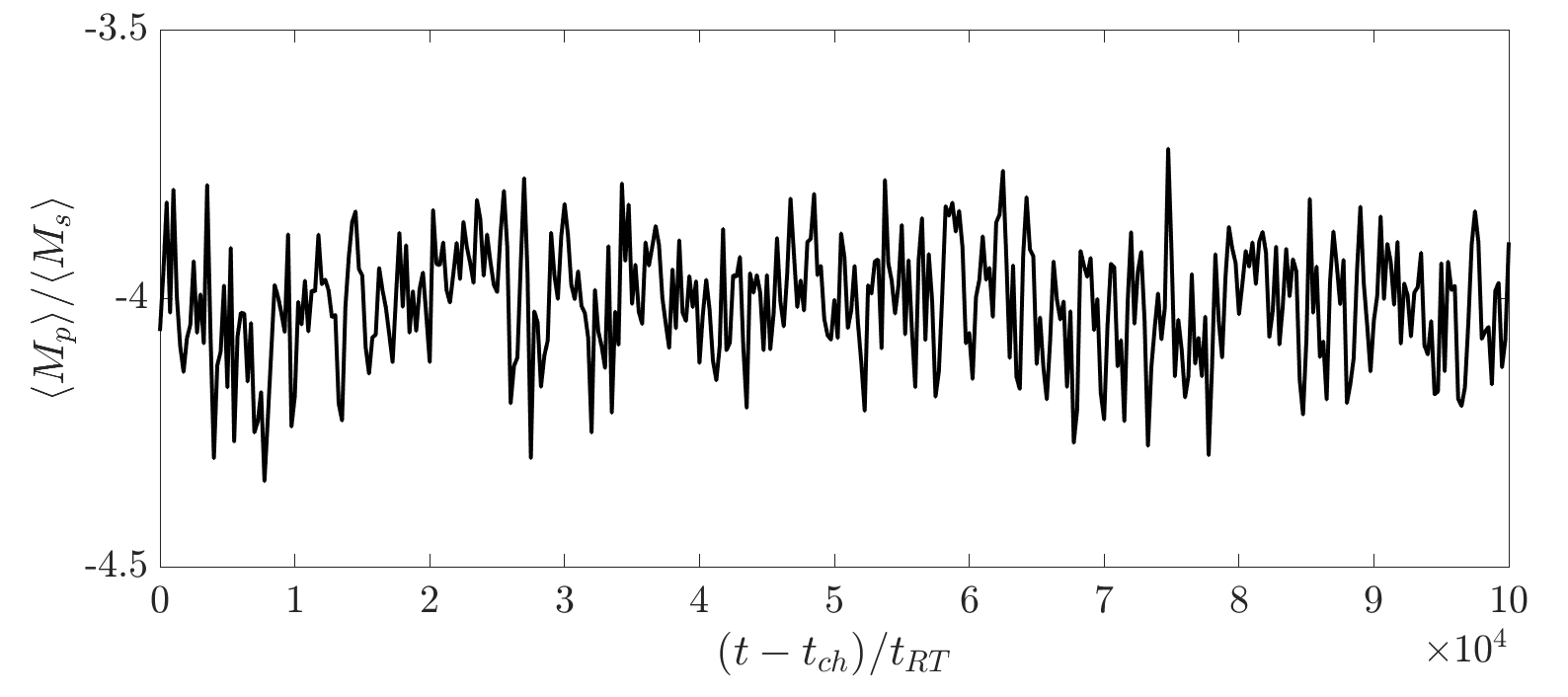}
    \caption{Plot of the ratio between ensemble average of the linear momentum for first and second harmonic over different times. $t_{ch}$ stand for the initial time when the system fully transits into a turbulent state. $t_{RT}$ denotes the round trip time $1/$FSR}
    \label{Fig:8}
\end{figure}

As a final remark, we will focus on the SHG dissipative breather gas, showing how the system reaches a statistically stationary equilibrium. This can be seen by measuring the ensemble average of the rhs of Eq.~\eqref{Eq:consM} over many realisations each characterised by a different initial random noise, see
Fig.~\ref{Fig:8}.
For this system, statistical stationarity implies that 
\begin{equation}
\frac{\langle M_p\rangle}{\langle M_s \rangle}=-\frac{\kappa_s}{\kappa_p},
\end{equation}
where $\langle \cdot\rangle$ stands for the ensemble average over many realisations. 
As pointed out in the appendix, if there is no walk-off the single SHG the solution is quiescent, $V=0$. This means that both $M_{p}$ and $M_{s}$ are equal to zero, as shown in Fig.~\ref{Fig:M}(a). The same idea can be applied to the case of statistical stationarity. If $\mathcal{U}$ is zero, then the ensemble averages $\langle M_{p}\rangle$ and $\langle M_{s}\rangle$ are also zero. On the other hand, when U is not zero, both $\langle M_{p}\rangle$ and $\langle M_{s}\rangle$ are non-zero, and their ratio is fixed by the dissipation value. In our system, where $\kappa_p/\kappa_s=4$, Fig.~\ref{Fig:8} shows how the ratio $M_p/M_s$ oscillates around the value -4. This observation suggests that despite the chaotic state, the first and second harmonics remain still locked together and move with a non-zero average velocity. We refer to this phenomenon as \emph{turbulence locking}.
\section{Conclusions}
We presented a study of the impact of walk-off on the stability of second harmonic generated soliton-comb in a realistic quadratic WGM microresonator. We showed that SHG breathers can be obtained in microresonator cavities away from the cascading regime. The soliton-breather transition can indeed be achieved away from the cascading regime due to the instability caused by the walk-off between the first and second harmonic light fields. Soliton instability is shown to be responsible also for the transition into a turbulent regime where a multitude of breathers coexists with dispersive waves.
A study of the PDF for the wave amplitude confirms the presence of a walk-off-induced SHG dissipative breather gas. Dissipative breather gas is then compared to its SHG dissipative soliton counterpart. Further analysis on the statistical average of the linear momentum showed the presence of locking between the turbulent state in the first and second harmonic.
Finally, a semi-analytical method is presented in order to estimate the velocity at which second harmonic induced combs move in the presence of walk-off. In parallel with this work, experimental observations of the SHG
breathers were reported in microresonators made of thin-film lithium-niobate~\cite{Lu2023Two-colourGeneration}.
\section*{Acknowledgements}
M.O. and A.V acknowledge the support provided by the Simon Collaboration on Wave Turbulence (Award ID 651741) and the Ministero dell’Università e della Ricerca under the PRIN program (Project No. 2020X4T57A).
D.V.S. acknowledges support from the Royal Society (SIF/R2/222029).
\appendix*
\section{Velocity Locking \label{App:A}} 
\begin{figure}
    \centering
    \includegraphics[width=0.48\textwidth]{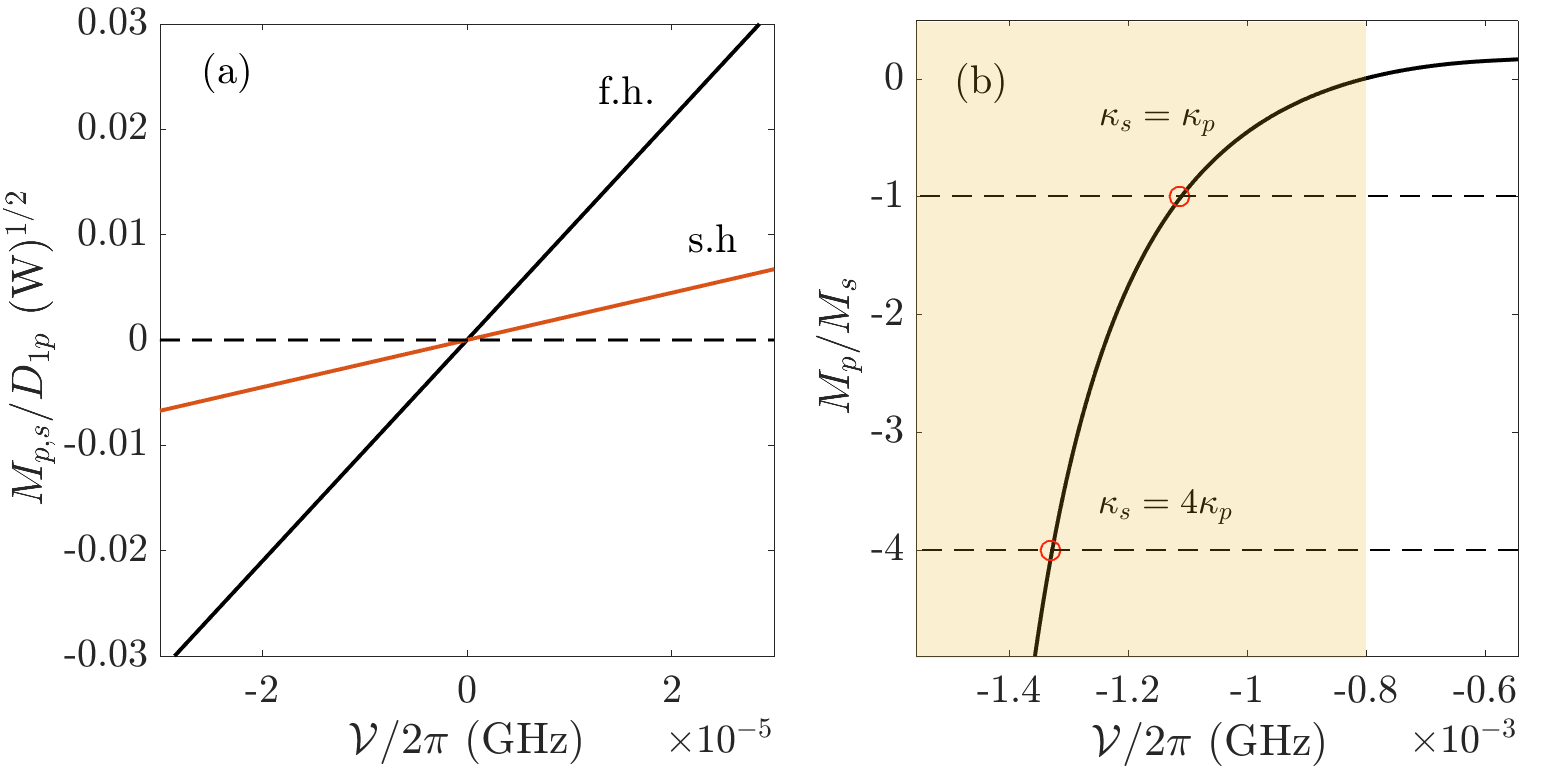}
    \caption{(a) Plot of linear momentum for first (black/dark gray line) and second harmonic (orange/light gray line) with respect to different soliton velocities in absence of walk-off.(b) Ratio of linear momenta vs soliton velocity for non zero walk-off, $\mathcal{U}/2\pi=5$ MHz. Red circle indicates numerically evaluated velocity selection for dissipation ratio $\kappa_s/\kappa_p=4$ and $\kappa_s/\kappa_p=1$. Detuning parameters and cw-laser power are the same as the one used in Fig.\ref{Fig:1}~(b).}
    \label{Fig:M}
\end{figure}

In this appendix we present an analysis on the velocity at which solitons are locked together. We start 
by evaluating numerically soliton solutions in the Hamiltonian limit.
For numerical reason it is easier to cast Eqs.\eqref{e1} and \eqref{e2} in dimensionless units by performing the following transformations: $A_{p,s}=\psi_{p,s}/\sqrt{\mathcal{W}} $, $\tau=D_{1p}t $, $d_{2p,2s}=D_{2p,2s}/D_{1p}$, $\Delta_{p,s}=\delta_{p,s}/D_{1p}$, $k_{p,s}=\kappa_{p,s}/D_{1p}$, $g_{p,s}=\gamma_{p,s}\sqrt{\mathcal{W}}/D_{1p}$ and $h=H/\sqrt{\mathcal{W}}$. In order to find soliton solutions moving with velocity $V=\mathcal{V}/D_{1p}$ in the presence of walk-off $U=\mathcal{U}/D_{1p}$ we will make use of a root-finding Newton-Raphson method to find stationary solution in a moving frame of reference, 
\begin{equation}
\theta\rightarrow \theta -(D_{1p}-\mathcal{V})t,  
\end{equation}
for the system of equations

\ba\label{eq:ad}
&&\nn \left(\Delta_p-iV\p_\ta-\tfrac{1}{2}d_{2p}\p^2_\ta\right)A_p-g_{p}A_s A_p^{*}+h=0\\
&&\left(\Delta_s-i(V-U)\p_\ta-\tfrac{1}{2}d_{2s}\p^2_\ta\right)A_s-g_{s} A_p^2 =0 .\ea 
We would like to stress that such a procedure can be easily applied in the Hamiltonian limit since for each value of walk-off parameter soliton solutions admit a continuous range of velocities. From Figure \ref{Fig:M}(a) and (b) it is possible to see how the linear momenta for each harmonic $M_{p,s}$ vary for different velocities. Even if we are dealing with an Hamiltonian limit, one can still extract further information concerning the locking velocity in the dissipative case by noticing that Eq.~\eqref{Eq:consM} forces moments to have opposite signs:
\begin{equation}\label{Eq:V_M}
    \frac{M_p(V)}{M_s(V)}=-\frac{\kappa_s}{\kappa_p}.
\end{equation}
From Figure \ref{Fig:M}(a) it clear that in the case of zero walk-off Eq.~\eqref{Eq:V_M} is never satisfied unless both $M_p$ and $M_s$ vanish, which is the case only when $V=0$. Differently, when $U\neq 0$, Eq.~\eqref{Eq:V_M} can be verified for a range of velocities, such as the one highlighted in yellow (light gray) in Fig.\ref{Fig:M}(b). From Eq.~\eqref{Eq:V_M} one can see how the relation between dissipation coefficients allows one to select a specific value of velocity locking. For example, Fig.\ref{Fig:M}(b) shows the velocity at which the soliton will move considering the value of the ratio $\kappa_s/\kappa_p$ equal to $1$ and $4$ respectively. The estimated velocity is remarkably close to the one obtained numerically making use of a velocity selective Newton-Raphson method, see red circles in Fig.\ref{Fig:M}(b). \\
We will now reintroduce dissipation in the system and present a semi-analytical method to estimate the value of the velocity locking. Such a method requires only the knowledge of the dissipative soliton solution $A^0_{p,s}=A_{p,s}(V=0,U=0)$ which is in the case of zero walk-off is easy to evaluate numerically since, as shown already, the locking velocity vanishes too.
Starting from the case when $U=0$ and $V=0$, one can assume that by introducing a small walk-off in the system, $U\sim\epsilon$, the solutions will modify accordingly, $A_{p,s}=A_{p,s}^{0}+\epsilon( A^{r}_{p,s}+i A^I_{p,s})$, resulting in first and second harmonic moving at non-zero locking velocity $V\sim\epsilon$. Considering Eqs.~\eqref{eq:ad} under such conditions and separating real and imaginary part, one can write the first order in $\epsilon$ as
\begin{equation}\label{Eq:eigen}
    \mathcal{M} \mathbf{A} + V\partial_\theta \mathbf{A}^{0}=U\partial_\theta \mathbf{A}^{(0)},
\end{equation}
where each column vectors contains real and imaginary part of the first and second harmonic components such that $\mathbf{A}=(A^r_p, A^i_p, A^r_s,A^i_s)^T$. Note that matrix $\mathcal{M}$ is given by
\begin{equation}\label{matrix}
\footnotesize
    \mathcal{M}=
    \begin{bmatrix}
-k_p -g_p (A^0_s)^i     && \mathcal{L}_p +g_p (A^0_s)^r &&g_p (A^0_p)^i  && -g_p (A^0_p)^r \\
-\mathcal{L}_p g_p (A_s^0)^r &&-k_p+ g_p (A^0_s)^i      && g_p (A^0_p)^r &&  g_p (A_p^0)^i \\
-2g_s(A^0_p)^i               &&-2g_s (A_p^0)^r               && -k_s          && \mathcal{L}_s       \\
 2g_s(A^0_p)^r               &&-2g_s (A_p^0)^i               && -\mathcal{L}_s       &&   -k_s        \\
\end{bmatrix}
\end{equation}
where we used the following definitions $\mathcal{L}_p\equiv(\Delta_p-1/2 d_{2p}\partial^2_\theta)$ and $\mathcal{L}_s\equiv(\Delta_s-1/2 d_{2s}\partial^2_\theta)$.
By noting that $\boldsymbol{\eta}\equiv\partial_\theta \mathbf{A}^{0}$ corresponds to the neutral eigenvector of $\mathcal{L}$, such that $\mathcal{M}\boldsymbol{\eta}=\lambda \boldsymbol{\eta}$ with $\lambda=0$, it is possible to derive the following equation~\cite{Skryabin2001WalkingSolitons}
\begin{equation}\label{Eq:semi_an_appndix}
    V=U\frac{\langle \boldsymbol{\phi} | \mathcal{P} \boldsymbol{\eta}\rangle}{\langle \boldsymbol{\phi} | \boldsymbol{\eta}\rangle}
\end{equation}
where $\mathcal{P} \boldsymbol{\eta}=(0, 0, \partial A^r_s,\partial A^i_s)^{T}$, $\langle \cdot | \cdot \rangle$ defines the scalar product and $\boldsymbol{\phi}$ corresponds to the neutral eigenvector of $\mathcal{M}^{\dagger}$. 
\begin{figure}
    \centering
    \includegraphics[width=0.48\textwidth]{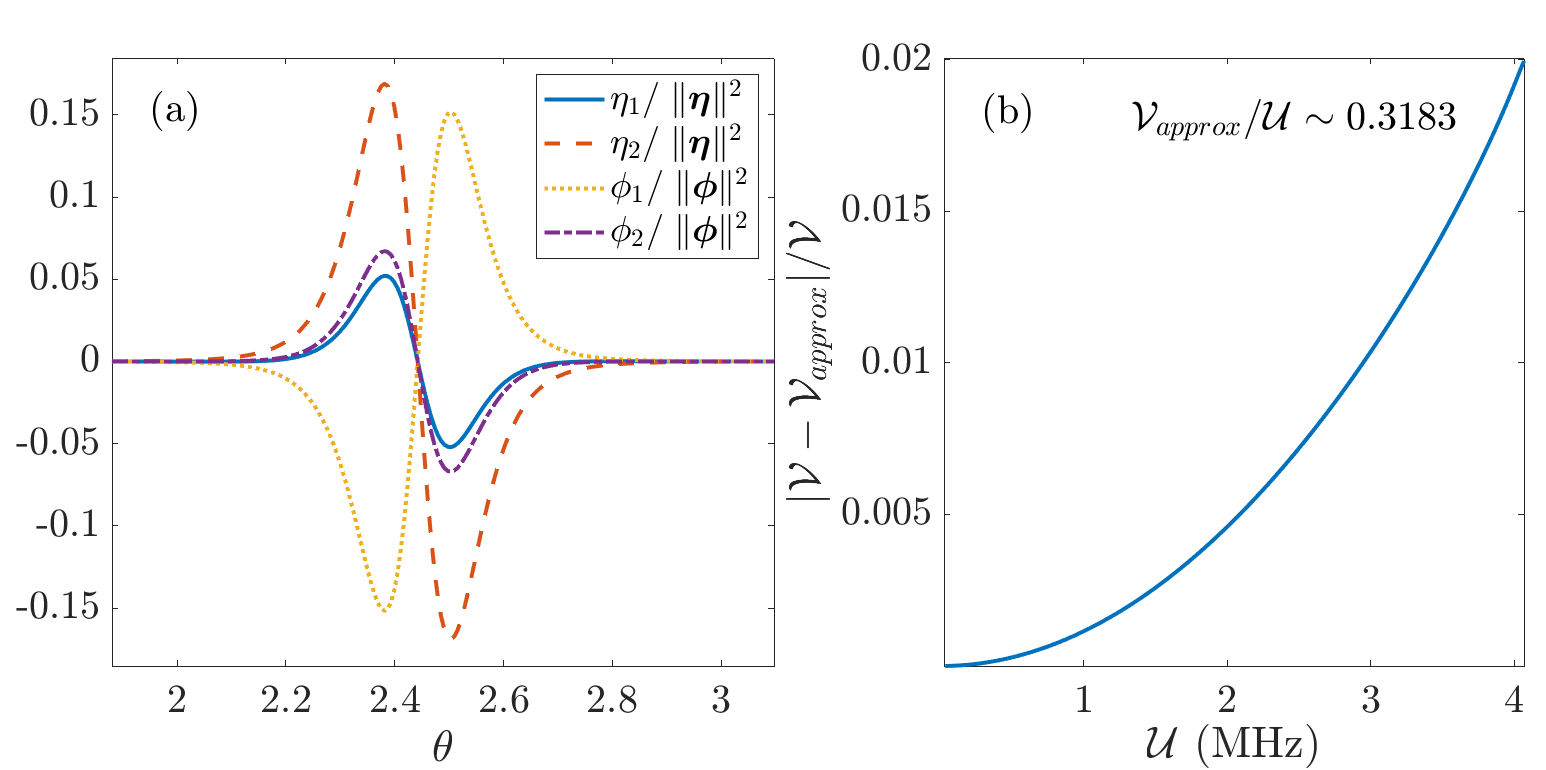}
    \caption{(a) First harmonic components of neutral eigenvectors of matrix.~\eqref{matrix} and its transpose. Each eigenvector is normalised to unity. (b) Relative error for analytical prediction for the soliton locking velocity}
    \label{Fig:3}
\end{figure}
Figure~\ref{Fig:3}(a) shows the first harmonic components of $\boldsymbol{\eta} $ and $\boldsymbol{\phi}$ respectively.
Despite the fact that Eq.~\eqref{Eq:semi_an_appndix} was derived in the limit of small walk-off, our semi-analytical formula provides excellent results for all possible values of walk-off where solitons exists, see Fig.~\ref{Fig:3} (b) where percent error is plotted in function of the value of walk-off parameter. 

\end{document}